# Understanding the limitation of Total Correlation Estimation Based on Mutual Information Bounds

Zihao Chen

**Abstract**

The total correlation(TC) is a crucial index to measure the correlation between marginal distribution in multidimensional random variables, and it is frequently applied as an inductive bias in representation learning. Previous research has shown that the TC value can be estimated using mutual information boundaries through decomposition. However, we found through theoretical derivation and qualitative experiments that due to the use of importance sampling in the decomposition process, the bias of TC value estimated based on MI bounds will be amplified when the proposal distribution in the sampling differs significantly from the target distribution. To reduce estimation bias issues, we propose a TC estimation correction model based on supervised learning, which uses the training iteration loss sequence of the TC estimator based on MI bounds as input features to output the true TC value. Experiments show that our proposed method can improve the accuracy of TC estimation and eliminate the variance generated by the TC estimation process.

## 1 Introduction

The estimation of independence between variables has a wide range of applications in many technical fields. Including bioinformatics[1][2][3], statistics[4][5][6], neuroscience[8],machine learning[12][13][14] etc. In machine learning, mutual information is often used as an important constraint item to measure the degree of disentanglement between variables, and common scenes including representation learning based on unsupervised disentangling[9][10][11], feature selection[15][16][17], generative-based disentangling model[12][18], clustering[19][20].

In statistics, the independence of paired variables is usually measured by mutual information (MI). Specifically, given a pair of variables $x$, $y$, the mutual information is represented by $I(\mathrm{x};\mathrm{y})$, which can be defined as:

$$I(\mathrm{x};\mathrm{y}) = E_{p(x,y)}\left[log\left(\frac{p(\mathrm{x},\mathrm{y})}{p(\mathrm{x})p(\mathrm{y})}\right)\right] \qquad (1)$$

where $x$ and $y$ are two one-dimensional variables, and $E$ is the expectation under the joint distribution $p(\mathrm{x},\mathrm{y})$.

To measure the correlation between numerous variables (more than two), we should utilize the total correlation (TC):

$$TC(Z) = E_{p(Z)}\left[log\left(\frac{p(\mathrm{Z})}{\prod_{i=1}^{n}p(z_i)}\right)\right] \qquad (2)$$

Where $Z$ is a multidimensional variable with $n(n>2)$ dimensions, and $TC(Z)$ obtains the total correlation among all dimensions of $Z$.

Since the exact computation of mutual information and total correlation is only tractable for discrete variables, or for a limited family of problems where the probability distributions are known [21]. Therefore, many methods have been proposed to estimate MI based on theoretical upper or lower bounds. For instance, MINE[22] maximizes an implicit function lower bound obtained by Jensen's inequality to estimate the mutual information between two random variables. InfoNCE[23] (Information Noise-Contrastive Estimation) is based on contrastive learning, which uses noise-contrastive estimation (Noise-Contrastive Estimation, NCE) as a bound to approximate mutual information, and uses neural networks to parameterize critic in NCE, so that it can be used more flexibly. The CLUB [24] algorithm employs unsupervised learning methods of logarithmic comparison and log-linear models to avoid overfitting and improve the accuracy of estimates. CLUB uses a technique called Contrastive Log-ratio Upper Bound to calculate the confidence interval and upper bound of the estimated TC value. This technique is based on the distribution of log comparison values and some statistical assumptions, which can effectively control the error and confidence of the estimate.

The same issue of an intractable computation arises when estimating the TC value of multidimensional variables without knowledge of the $p(x)$ distribution. In representation learning, a straightforward prior distribution assumption of $p(x)$ is frequently made to handle this issue[11][25]. In [26], a variable $y$ is added, it is assumed that all $x_i$ are orthogonal to $y$, and a theoretical upper bound of TC is derived. Whereas in [27], two total correlation decomposition paths are proposed, which decompose the TC of multidimensional variables into the sum of MI estimates of multiple pairwise variables. This approach can apply TC estimation to a broader scenario and doesn't call for any additional presumptions.

In this paper, the accuracy of TC estimation based on MI decomposition is theoretically shown to be influenced by the difference between the proposal distribution and the target distribution of MI items based on importance sampling. Through empirical experiments under different MI estimation methods, we qualitatively discovered that the impact of this bias is amplified as the true TC rises. Since there is some link between the true TC and the performance of the estimator during the estimate process, we propose employing a parameter learnable model built on neural networks as a TC estimation bias corrector. By learning the loss value sampling sequence throughout the TC estimator, a more precise TC value is inferred based on the supervised learning method. We performed experiments using a range of TC estimation methods and discovered that this approach is workable for the majority of MI-bounds-based TC estimators, significantly improves the estimation accuracy of TC, and can eliminate the variance in the estimation process.

## 2 Deviation Analysis of TC Estimation Based on MI Bounds

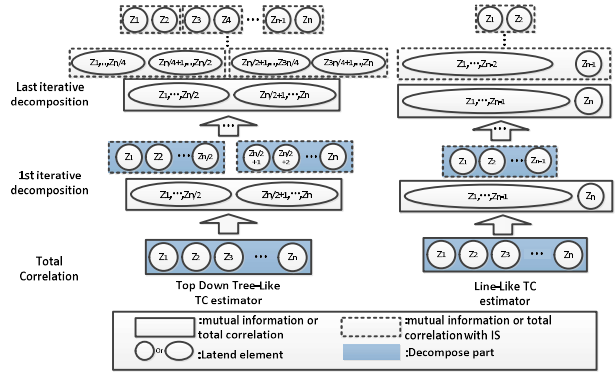

Figure 1. Two iterative decomposition paths of TC. Suppose that latent variable z has n dimensions: the circle represents marginal distribution of a variable; the ellipse represents joint distribution of several variables; the blue color area represents the term to be decomposed from the current to the next iteration; the dotted line in the box refers to the MI item based on importance sampling.

In Cheng's work[27], the total correlation and mutual information are connected, and two paths for decomposing the total correlation are proposed:

Assuming the latent variable $z$ has $n$ dimensions. Then, $n$ can be divided into two sets: $Z_A := (z_{k1}, z_{k2}, z_{k3} \ldots, z_{km})$, $Z_{\bar{A}} := (z_{v1}, z_{v2}, z_{v3} \ldots, z_{vt})$, making $Z = Z_A \cup Z_{\bar{A}}$, $Z_A \cap Z_{\bar{A}} = \emptyset$. The Top-Down Tree-like path decomposition is written as follows:

$$TC(Z) = E_{p(Z)}\left[\log\left(\frac{p(Z_A)}{p(z_{v1})p(z_{v2})\ldots p(z_{vt})}\right)\right] + E_{p(Z)}\left[\log\left(\frac{p(Z_{\bar{A}})}{p(z_{v1})p(z_{v2})\ldots p(z_{vt})}\right)\right] +$$

$$E_{p(Z)}\left[\log\left(\frac{p(Z)}{p(Z_A)\cdot p(Z_{\bar{A}})}\right)\right] = TC(Z_A) + TC(Z_{\bar{A}}) + I(Z_A \; ; \; Z_{\bar{A}}). \tag{3}$$

Similarly, The Line-Like path:

$$TC(Z_{1:n}) = E_{p(Z)}\left[\log\left(\frac{p(Z)}{p(z_1,z_2,\ldots z_{n-1})p(z_n)}\right)\right] + E_{p(Z)}\left[\log\left(\frac{p(z_1,z_2,\ldots z_{n-1})}{p(z_1)p(z_2)\ldots p(z_{n-1})}\right)\right] = I(Z_{1:n-1} \; ; \; z_n) +$$

$$TC(Z_{1:n-1}) = \sum_{i=1}^{n-1} I(Z_{1:i} \; ; \; Z_{i+1}), \tag{4}$$

where $min(n, i+2)$ represents the smaller value among $i+2$ and $n$.

These two paths break down the immeasurable total correlation into several pairwise MI groups. so that we can derive total correlation estimators based on the previous mutual information bounds.

However, we re-derived the total correlation decomposition process and discovered that some items were not mutual information according to information theory but rather mutual information based on importance sampling (for a detailed derivation, see the supplementary material A). For instance, $E_{p(Z)}\left[\log\left(\frac{p(Z_A)}{p(z_{v1})p(z_{v2})\ldots p(z_{vt})}\right)\right]$ in formula (3) can be further deduced as $E_{p(Z_A)}[p(Z_{\bar{A}}|Z_A) \cdot \log(\frac{p(Z_A)}{p(z_{k1})p(z_{k2})\ldots p(z_{km})})]$. We regard $p(Z_{\bar{A}}|Z_A)$ in the above term as the important weight $w$. From the fact that $w$ is equal to 1 when $Z_A$ and $Z_{\bar{A}}$ are independent. It can be inferred that $E_{p(Z_A)}[p(Z_{\bar{A}}|Z_A) \cdot \log(\frac{p(Z_A)}{p(z_{k1})p(z_{k2})\ldots p(z_{km})})]$ should be seen as the total correlation of $Z_A$, but it needs to be obtained through importance sampling when $Z_A$ and $Z_{\bar{A}}$ are independent. We will write it as $TC(Z_A)|_{q(Z)=q(Z_A)\cdot q(Z_{\bar{A}})}$. Similarly, $E_{p(Z_{\bar{A}})}[p(Z_A|Z_{\bar{A}}) \cdot \log(\frac{p(Z_{\bar{A}})}{p(z_{v1})p(z_{v2})\ldots p(z_{vt})})]$ is also the total correlation of $Z_{\bar{A}}$ estimated through sampling when the target distribution satisfies that $Z_A$ and $Z_{\bar{A}}$ are independent of each other. We will write it as $TC(Z_{\bar{A}})|_{q(Z)=q(Z_A)\cdot q(Z_{\bar{A}})}$.

After modifying all similar items in the formula (3), Top-Down Tree-like path decomposition should be revised as follow:

$$TC(Z) = TC(Z_A)|_{q(Z)=q(Z_A)\cdot q(Z_{\bar{A}})} + TC(Z_{\bar{A}})|_{q(Z)=q(Z_A)\cdot q(Z_{\bar{A}})} + I(Z_A \ ; \ Z_{\bar{A}}). \quad (5)$$

Similar to Top-Down Tree-like path decomposition, the revised decomposition of the Line-Like path should be as follows:

$$TC(Z_{1:n}) = I(Z_{1:n-1} \ ; \ Z_n) + \sum_{i=1}^{n-2} I(Z_{1:i} \ ; \ Z_{i+1})|_{q(Z)=q(z_{\min(n,i+2):n})\cdot q(z_{1:i+1})}, \quad (6)$$

where $min(n, i+2)$ represents the smaller value among $i+2$ and $n$.

Refer to Fig 1 for the decomposition process of the two revised decomposition paths

Consequently, the following interesting conclusions can be drawn from the above discovery: First, to avoid significant deviations in the sampling results, the general trend of the proposal distribution in importance sampling should be consistent with the target distribution. The two distributions cannot have significantly different distribution characteristics, otherwise, the sampling estimation results will have a large deviation [33]. As a result, in real estimation scenarios, if this distribution difference is not limited, then this will be a "hidden danger" of bias in TC estimation. In the experimental section, we will show that when the true TC value is relatively high, the deviation of the importance sampling procedure has a greater effect on the precision of the TC estimate.

## 3 TC estimation corrector

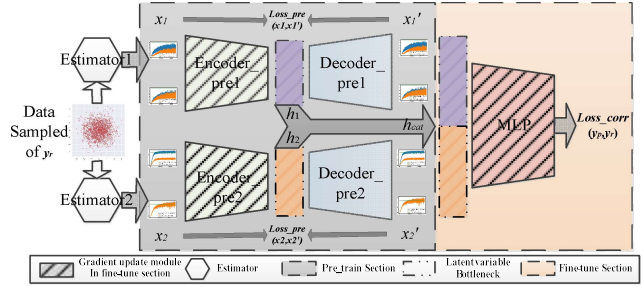

Figure 2. The figure depicts the training of a multi-input corrector. At the pre-training parts, Encoder_pre1 and Encoder_pre2 respectively receive the loss value sequences x1 and x2 generated by different estimators during the estimation process, and extract the features in the sequences into a low-dimensional latent spaces h1 and h2 through the under-complete autoencoder and splicing them into hcat, and then input hcat to the corrector for supervised training of the fine-tune part, finally use Loss_Corr loss backpropagation to adjust the trainable parameters in MLP and encoder_pre1 and encoder_pre2.

Because the total correlation estimate technique is based on MI Bounds, the trade-off between estimation bias and variance is as challenging as it is for mutual information estimation. Existing studies have proved that the variance of certain estimators could grow exponentially with the ground truth MI [28]. Moreover, from the theoretical derivation in the previous chapter, we also found that as the true TC increases, the estimation bias may be amplified. Since there is some association between the estimator's performance and the true TC value. To more correctly estimate the TC, we were inspired to create a simple-enough parameter corrector, which aims to learn the characteristic information of the bias from the estimation process of the estimator and try to correct the estimation bias.

Begin from collecting the labeled dataset, it is challenging to identify the true distribution of sampled data in a real application, but we can simply create the covariance matrix of multi-dimensional variables to precisely obtain its total correlation, and then data sampling and TC estimation are applied to the distribution of these known true TC to generate an enormous number of loss value sequences with labels (true TC values), which can be utilized for corrector training. While applying the model, we typically initiate with a large number of samples from an unknown distribution. Then input these samples into the estimator to obtain the loss value sequence, after which we can employ the trained corrector to output the true TC value.

Specifically, we design a neural network-based corrector behind the TC estimator. The corrector is composed of (one or more) under-complete autoencoders and an MLP parameter learner. The structure of the corrector refers to Fig 2. The training process consists of two stages: the pre-training stage and the fine-tuning stage. During the pre-training stage, the under-complete autoencoder is trained. Encoder_pre receives the sequence of loss values $x_1$ produced by the estimator, and Decoder_pre outputs the reconstruction sequence $x_1'$. The features of the loss sequence can be captured by the latent variables at the bottleneck of under-complete autoencoder during pre-training, improving the convergence stability of the subsequent fine-tuning stage[29]. Functionally speaking, the under-complete autoencoder compresses and transfers the estimator's training process feature information to a low-dimensional space, enabling us to combine the estimate process features of various estimators for fine-tuning. As a result, the model's ability to learn features from multiple perspectives during the fine-tuning stage enhances the total quantity of information that the corrector may capture. We refer to such a corrector fed by multiple estimators as a multi-input corrector. The bottleneck latent variable from the pre-training, $h_{cat}$, is fed into an MLP parameter learner in the fine-tuning stage, and the output is a single-valued TC estimation, $y_p$, which is used to calculate the loss function, $Loss\_corr(y_p, y_r)$. Finally, back-propagating gradients modify the trainable parameters in the corrector (Refer to Fig 2 for specific parts that need to be modified).

The loss function optimized for the entire training can be simplified to the following objective:

$$Loss = \sum_{i=1}^{n} Loss\_pre(x_i, x_i) + Loss\_corr(y_p, y_r), \qquad (7)$$

Where $Loss\_pre$ represents the loss function in the pre-training stage, $n$ is the total number of MI estimators, and $Loss\_corr$ represents the loss function in the fine-tuning stage.

## 4 Experiment

### 4.1 Qualitative Analysis of Estimation bias

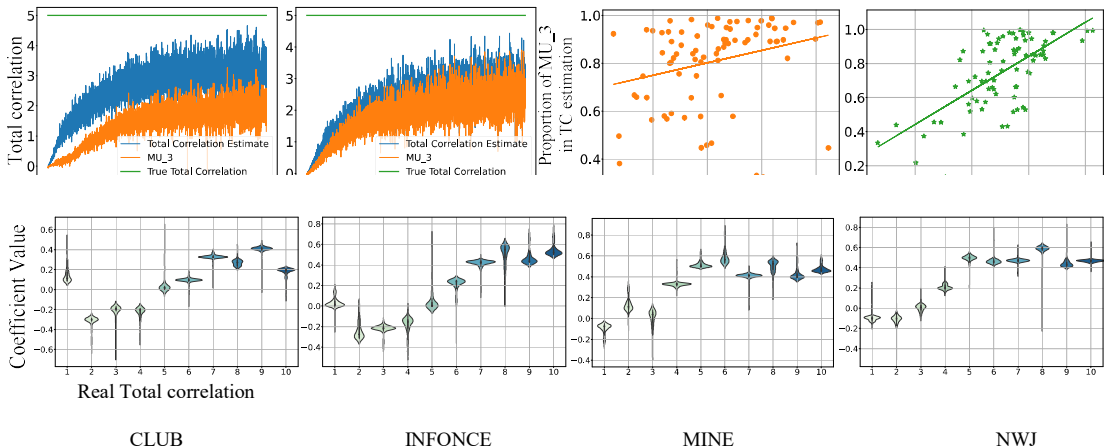

Figure 4. The figure above shows the Spearman correlation between the proportion of the third mutual information in the TC estimation and the estimated absolute value error under different true total correlations. CLUB, INFONCE, MINE, and NWJ are the MI estimate methods utilized in the figure, from left to right. The figure shows that as the true total correlation increases, so does the positive correlation between the estimation error and the size of the third term. This correlation expresses weak in the CLUB method while relatively strong in other methods.

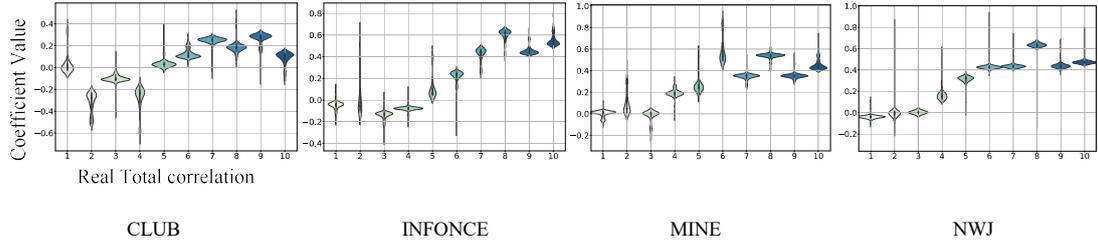

CLUB  INFONCE  MINE  NWJ

Figure 5. Same as graph n, but with the Pearson correlation used in this experiment. The effect of the experiment is almost consistent with the Spearman correlation experiment.

Through analysis of different paths of total correlation decomposition in section 2, we find that all paths have partial estimates based on importance sampling. In specific scenarios, the estimated bias may become greater if the proposed distribution is too different from the target distribution. In this section, we empirically demonstrate this conjecture.

Assuming that the data $z$ is a 4-dimensional multidimensional variable that satisfy Gaussian distribution $(Z_1, Z_2, Z_3, Z_4) \sim N(0, \Sigma)$. $\Sigma$ is a symmetric covariance matrix. The true TC value of $z$ can be obtained by calculating $-logDet(\Sigma)$.

The revised 4-dimensional top-Down Tree-like path decomposition can be written as follows:

$$TC(z) = I(Z_1\ ;\ Z_2)|_{q(Z)=q(Z_{1,2})\cdot q(Z_{3,4})} + I(Z_3\ ;\ Z_4)|_{q(Z)=q(Z_{1,2})\cdot q(Z_{3,4})} + I(Z_{1,2}\ ;\ Z_{3,4}). \quad (8)$$

We need to notice, the first two terms of objective (8) are based on the mutual information estimation of importance sampling with target distribution obeying that $q(Z_{1,2})$ and $q(Z_{3,4})$ are independent of each other. Hence, the third term $I(Z_{1,2}\ ;\ Z_{3,4})$ can be regarded as the constraint item of the difference between the target and proposed distribution in the first two terms. We design experiments to observe the relevance between the relative size of third-term mutual information and TC estimation accuracy.

In the experiment, MINE, infoNCE, CLUB, and NWJ[30] algorithms are used as TC estimators. The true TC values in the experiments range from 1 to 10, with 5000 iterations in a single training round. We randomly generate 100 different eligible covariance matrices for each true TC value. Randomly group 100 eligible covariance matrices into 10 groups to train and estimate with four algorithms. We record the estimated TC value and the proportion of the third item mutual information in the estimated TC value during the estimation and try to find the connection between the third term and the estimation bias, as a sampling example shown in Fig. 3.

Pearson correlation [31] needs to assume that the original data satisfies the normal distribution, while the real distribution of the samples in the experiment is unknown. Whereas Spearman correlation [32] does not need to assume the data distribution and is more friendly to outlier samples. Therefore, we conducted the same experiment based on the Spearman correlation and Pearson correlation respectively. It can be seen from Fig. 4 and Fig. 5 that the results of the two groups of experiments have strong consistency.

When the true TC is greater than 5, the Pearson and Spearman correlation is about 0.5~0.6, indicating that the size of the third item has a relatively strong negative correlation with the precision of the TC estimation. The greater the difference between the proposed and target distribution, the greater the third item and the less accurate the TC estimation. Specifically, when the TC is sufficiently small (TC<5), the third-term mutual information value must be smaller, so the proportion of the third item in the TC estimation has less of an effect on the accuracy.

According to experimental findings, the bias impact of the difference between the proposed distribution and the target distribution of important sampling items in decomposition on the accuracy of TC estimation grows as the true TC value rises.

## 4.2 Performance experiment of corrector
### 4.2.1 Experimental settings and CORR_Dataset

In this section, we will demonstrate the superiority of the corrector in terms of TC estimation accuracy and variance through experiments. Finally, we will also show the robustness of the corrector model.

The TC value of the 4-dimensional variable requested to be estimated in the experiment. To begin, we build enough 4-dimensional covariance matrices at random to ensure that the true TC value covers the range of 0~10 flatly. The samples generated by the distributions carrying these covariances are then estimated using a list of MI estimation algorithms, with the sequence of loss values produced by each algorithm getting recorded. To undertake TC estimate experiments, we employ the MINE, infoNCE, NWJ, and CLUB algorithms, as well as correctors based on these algorithms.

To train the corrector in a supervised manner, we collected loss sequences from the estimation process of the four estimators for different data distributions. Specifically, each algorithm performs 3k rounds of training for the estimation process of each distribution data. We interpolate and sample loss sequences every 100 iterations to obtain 30 loss value sequences. 200 covariance matrices are randomly generated at each gap of 1 in TC value 0~10. We collected a total of 8k sets of labeled loss sequences generated by four methods for 2k different distribution matrices for corrector supervised learning in the experiment. We named this dataset CORR_ Dataset.

### 4.2.3 Impact of input perspective

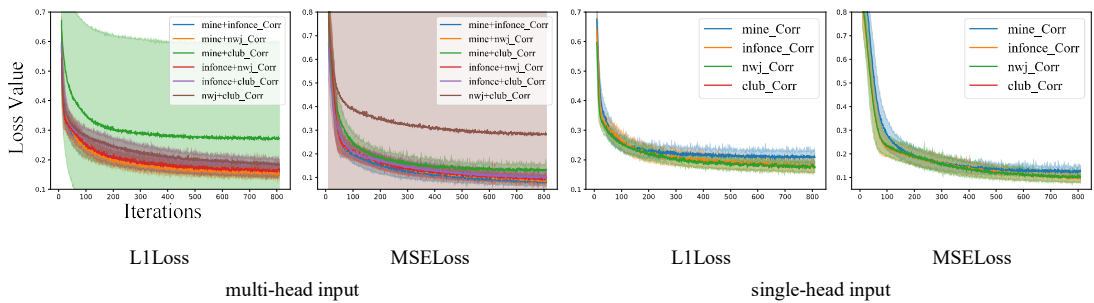

L1Loss     MSELoss     L1Loss     MSELoss

multi-head input     single-head input

Figure 6. We experimented on correctors based on 4 kinds of MI estimates. In each of the above figures, the ordinate is the loss value, and the abscissa is the number of iterations. The transparent filled part is the loss interval of 90% confidence in the repeated training process. Among them, the single-head input corrector based on CLUB estimation cannot converge to the error region of other methods, so it does not appear in the two figures on the right. Among the single-head input correctors, the corrector based on infoNCE and NWJ methods has the smallest test error. However, the combination of CLUB in the multi-input corrector will lead to an increase in error and a large training variance. The best effect is the multi-head corrector based on the combination of MINE+infoNCE and MINE+NWJ methods. Overall, the correctors based on multi-head inputs are able to achieve smaller TC estimation errors.

The first set experiments compared the test error of multi-head input correctors versus single-head input correctors during training. We randomly split the training set into 3:7 (3 test sets and 7 training sets) in the experiment. In the multi-input experiments, six different multi-input correctors were composed using four different estimation algorithms. The two images on the left side of Fig 6 illustrate the multi-head input corrector's test error. The multi-head input corrector with MINE+infoNCE has the best testing error under MSELoss, while the best under L1Loss is the corrector with MINE+NWJ combination.

The two images on the right are experiments with single-head input correctors, and the corrector based on infoNCE and NWJ methods has the smallest testing error. Among them, the corrector based on the CLUB estimation method is relatively ineffective and cannot converge to the test error range of other correctors. In contrast, multi-input correctors generally achieve smaller test errors than single-input correctors.

### 4.2.4 Estimation accuracy and variance of the corrector

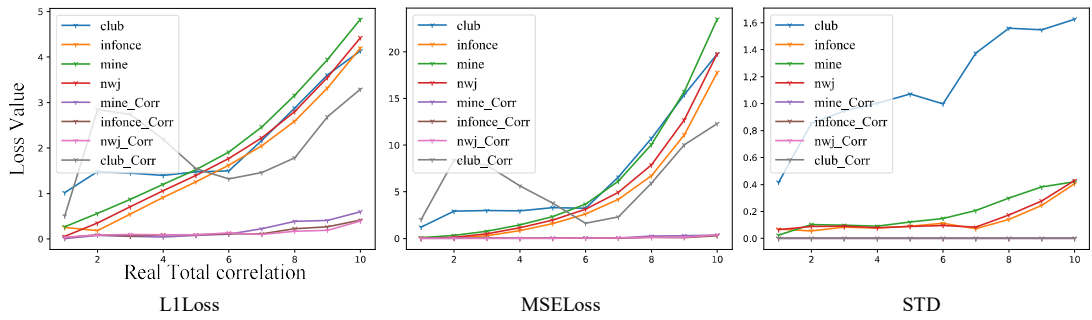

L1Loss  MSELoss  STD

Figure 7. Under the real total correlation of 1~10, observe the total correlation estimation error and variance of the 4-dimensional variables based on different estimators and correctors. It can be seen from the above figure that based on the four pure estimators, the estimation error will increase significantly with the increase of the true total correlation. Among the corrector methods, except for the mehod based on the CLUB , which has a large error (but still smaller than any pure estimator), the errors of all the methods are very small, and the increase of TC has little effect on the accuracy of these correctors. In the variance experiment on the right, the pure estimator also shows a trend of increasing the estimated variance as the true TC increases, and the variance of the CLUB method increases most obviously. However, the method based on the corrector has no estimation variance due to the design of a single output neuron in the model.

The second set of experiments evaluates the estimator's and corrector's estimation bias performance under varied true TC values. In the experiment, the corrector's training set is randomly split into testing set and training set at 4:6 ratio. For experimental accuracy, the corrector performs 30 times independent training and testing and finally presents the average test error. From Fig 7, it is evident that the estimated bias and variance of the pure estimator grow as the true TC value increases. In contrast, most correctors'(Only the CLUB-based corrector has a larger estimation bias than the pure estimator when the TC is small) estimate bias will only marginally rise as TC increases. In the variance estimation experiment, it is obvious that the variance of the pure estimator grows constantly as TC increases, and the magnitude of the CLUB method is the steepest. The corrector has no estimation variance because it is based on single-neuron output.

From the above two groups of experiments, it can be concluded that in most cases, the corrector can greatly improve the TC estimation accuracy of the model compared with the pure estimator, and eliminate the estimation variance. On this basis, the corrector with multi-head input can further reduce the test error during training compared with the corrector with single-head input.

### 4.2.5 Model robustness under imperfect training set

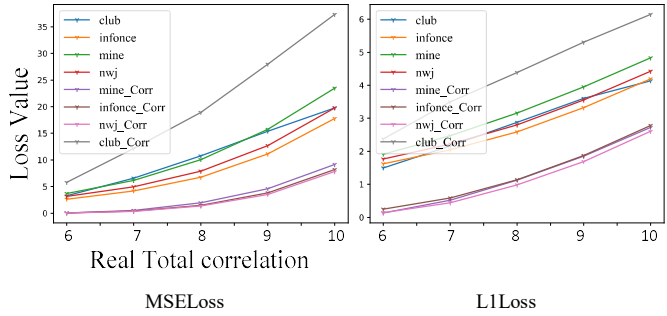

Figure 8. To evaluate the model's robustness in the event of an imperfect training set, the training set of the corrector masks all samples with TC true values ranging from 6 to 10. Under the two loss functions, most correctors (except CLUB-based correctors) can still guarantee lower estimation errors than pure estimators. Specifically, the one with the smallest generalization error under both two loss functions is the corrector of the NWJ method

The estimation method proposed in this paper is a training data-driven model. When applying the corrector, it is necessary to assume that the training set and the validation set satisfy independent and identical distribution. Nevertheless, we cannot ensure that the distribution characteristics produced in the training set would fully cover the actual distribution faced by the model in a real-world application. Therefore, we are more concerned about the model's estimation accuracy when faced with unseen distributions.

In this part, we simulate and illustrate the corrector model's robustness given an imperfect training set. To test the model's generalization performance under missing data, we divide the training set with actual TC values ranging from 0 to 10 into two groups: those with TC values between 0 and 5 are used as the training set, while those with TC values between 6 and 10 are used as verification set.

The verification error based on MSELoss is shown on the left side of Fig 8, while the verification set error based on L1Loss is shown on the right. Apart from the corrector based on the CLUB estimator, all correctors can reduce estimation bias when compared to pure estimators. This experiment demonstrates that even with imperfect training data, most correctors can maintain good robustness.

### 5 discussion

We uncover sources of bias in total correlation estimates based on MI bounds decomposition. It is theoretically deduced that this deviation is related to the use of importance sampling in the decomposition process, and it is qualitatively demonstrated that the impact of this deviation will be amplified as the true TC increases. This will provide a theoretical basis for error analysis of deep learning methods based on TC estimation. Inspired by this, we propose a TC estimation bias correction model based on supervised learning. Experiments prove that this correction model can enhance TC estimation accuracy while maintaining robustness in the case of imperfect training data. Furthermore, this method avoids the issue of estimation variance being increased as the true TC value grows during the estimation phase. In practice, if TC is unconstrained (TC value may be relatively large), the method proposed in this paper is unquestionably more suitable.